\documentclass[apj]{emulateapj}
\usepackage{mathptmx}

\usepackage{verbatim}
\usepackage{graphicx}
\usepackage{epstopdf}
\usepackage{amsmath}
\usepackage{textcomp}
\usepackage{latexsym}
\usepackage{multirow}
\usepackage{wasysym}

\newcommand{\gtsim}{\raisebox{-.5ex}{$\;\stackrel{>}{\sim}\;$}}
\newcommand{\kms}{\ifmmode {\rm km\ s}^{-1} \else km s$^{-1}$\fi}
\newcommand{\lledd}{$L/L_{\rm Edd}$}

\newcommand{\et}{et al.\ }
\newcommand{\xray}{\hbox{X-ray}}

\newcommand{\spitzer}{{\sl Spitzer}}
\newcommand{\lya}{Ly$\alpha$}
\newcommand{\hb}{H$\beta$}
\newcommand{\mopex}{{\sc mopex}}
\newcommand{\apex}{{\sc apex}}

\journalinfo{The Astrophysical Journal, ???:???, 2011 ??? ??}
\slugcomment{Received 2011 June 2; accepted 2011 September 23}

\shortauthors{LANE ET AL.}
\shorttitle{UV-TO-MID-IR SED OF WLQS}
\begin{document}
%%%
%%%
%%%
\title{The Ultraviolet-to-Mid-Infrared Spectral Energy Distribution of \\ Weak Emission Line Quasars}

\author{
Ryan~A.~Lane,\altaffilmark{1}
Ohad~Shemmer,\altaffilmark{1}
Aleksandar~M.~Diamond-Stanic,\altaffilmark{2,3}
Xiaohui~Fan,\altaffilmark{4}
Scott~F.~Anderson,\altaffilmark{5} 
W.~N.~Brandt,\altaffilmark{6,7}
Richard~M.~Plotkin,\altaffilmark{8}
Gordon~T.~Richards,\altaffilmark{9}
Donald~P.~Schneider,\altaffilmark{6}
and Michael~A.~Strauss\altaffilmark{10}
}
\altaffiltext{1}
		   {Department of Physics, University of
                      North Texas, Denton, TX 76203, USA; RyanLane@my.unt.edu, ohad@unt.edu}
\altaffiltext{2}
                      {Center for Astrophysics and Space Sciences,
                      University of California, San Diego, La Jolla, CA, 92093, USA}
\altaffiltext{3}
                      {Center for Galaxy Evolution Fellow}
\altaffiltext{4}
		   {Steward Observatory, University of Arizona, 933 North Cherry
                      Avenue, Tucson, AZ 85721, USA}
\altaffiltext{5}
                      {Department of Astronomy, University of Washington, Box 351580,
                      Seattle, WA 98195, USA}
\altaffiltext{6}
		   {Department of Astronomy \& Astrophysics, The Pennsylvania State
                      University, University Park, PA 16802, USA}
\altaffiltext{7}
                     {Institute for Gravitation and the Cosmos, The Pennsylvania State
                     University, University Park, PA 16802, USA}
\altaffiltext{8}
		  {Astronomical Institute `Anton Pannekoek', University of Amsterdam,
                     Science Park 904, 1098 XH, Amsterdam, The Netherlands}
\altaffiltext{9}
		   {Department of Physics, Drexel University, 3141 Chestnut Street,
                      Philadelphia, PA 19104, USA}
\altaffiltext{10}
 		   {Princeton University Observatory, Peyton Hall, Princeton,
                      NJ 08544, USA}

%%%
%%%
%%%
\begin{abstract}
We present \spitzer\ Space Telescope photometry of 18 Sloan Digital Sky Survey (SDSS) quasars at \hbox{$2.7 \leq z \leq 5.9$} which have weak or undetectable high-ionization emission lines in their rest-frame ultraviolet (UV) spectra (hereafter weak-lined quasars, or WLQs).
The \spitzer\ data are combined with SDSS spectra and ground-based, near-infrared (IR) photometry of these sources to produce a large inventory of spectral energy distributions (SEDs) of WLQs across the rest-frame \hbox{$\sim0.1-5~\mu$m} spectral band.
The SEDs of our sources are inconsistent with those of BL Lacertae objects which are dominated by synchrotron emission due to a jet aligned close to our line-of-sight, but are consistent with the SED of ordinary quasars with similar luminosities and redshifts that exhibit a near-to-mid-IR `bump', characteristic of hot dust emission. This indicates that broad emission lines in WLQs are intrinsically weak, rather than suffering continuum dilution from a jet, and that such sources cannot be selected efficiently from traditional photometric surveys.
\end{abstract}
%%%
%%%
%%%

\keywords{galaxies: active -- galaxies: nuclei -- infrared: galaxies --
  quasars: general -- quasars: emission lines}

%%%
%%%
%%%
\section{Introduction}
\label{sec:introduction}
%%%

The spectra of most optically-selected quasars are marked by strong, broad emission lines in the rest-frame ultraviolet (UV) band.
However, the Sloan Digital Sky Survey (SDSS; York \et  2000) has, so far, discovered $\sim80$ quasars at $z\apprge2.2$ with weak or undetectable rest-frame UV emission lines (hereafter weak-lined quasars, or WLQs), starting with the prototype
SDSS~J153259.96$-$003944.1 (Fan \et 1999; see also Anderson \et 2001; Collinge \et 2005; Diamond-Stanic \et 2009, hereafter DS09; Plotkin \et 2010a).
Such WLQs can be defined as sources having rest-frame equivalent widths (EWs) of $\leq15.4$\,\AA\ for the \lya$+$\ion{N}{5} emission complex; this threshold represents the $3\,\sigma$ limit at the low-EW end of the lognormal distribution of EW[\lya$+$\ion{N}{5}]  in $z>3$ SDSS quasars (DS09).
Due to their largely featureless spectra, the redshifts of these sources can only be determined reliably from the onset of the \lya\ forest (i.e., \hbox{$\lambda_{\rm rest}  \sim 1200$\,\AA}). The discovery of WLQs only at $z\apprge2.2$ is clearly a selection effect,
as the onset of the \lya\ forest emerges in the SDSS spectral range at that redshift.
Indeed, there are indications for a population of lower-redshift sources with similar characteristics  to the high-redshift WLQ population (e.g., McDowell \et 1995; Leighly \et 2007a; Hryniewicz \et\ 2010, Plotkin et al. 2010b).

Since their discovery, a number of scenarios have been proposed to explain the properties of these remarkable objects, although many have been subsequently shown to be unlikely.
For example, WLQs are unlikely to be broad absorption line (BAL) quasars, as their spectra do not show obvious
broad \ion{C}{4} absorption features.
In addition, Shemmer \et (2006; hereafter S06) show that WLQs are not likely to be active galactic nuclei (AGNs) where
emission lines have
been obscured by dust and that they are not normal galaxies with luminosities that have been amplified by gravitational lensing. 
Spectroscopic monitoring of
four WLQs further suggests that it is unlikely that the weakness
of their lines can be explained by microlensing that temporarily
and preferentially amplifies the continuum relative to the broad
emission lines (DS09).
Instead, S06 proposed that WLQs are either the long-sought, high-redshift counterparts to BL Lacertae
(BL Lac) objects (e.g., Stocke \& Perrenod 1981) or that they are `ordinary' unbeamed, unobscured quasars with extreme properties
such as a deficit of line-emitting gas in the
broad emission line region (BELR)
or an extremely high accretion rate (see also Fan \et 1999).
It has also been speculated that WLQs may represent an early or transitional
phase in quasar evolution where quasar activity just recently `switched on'
(Hryniewicz \et 2010; see also Liu \& Zhang 2011).

Classical BL Lac objects are a rare class of AGNs with nearly featureless spectra that are radio loud, \xray\ strong, polarized, and highly variable. All these characteristics are attributed to the fact that BL Lacs are AGNs viewed along their powerful, narrow jets (e.g., Urry \& Padovani 1995).
In particular, the nearly featureless spectrum of BL Lacs is a consequence of relativistically-boosted synchrotron emission from the jet that overwhelms the characteristic AGN emission lines.
In contrast, WLQs are, at most, radio- and \xray-intermediate, exhibiting only minimal overlap with the radio- and \xray-weak tails of the BL Lac population
(e.g., Shemmer \et 2009, hereafter S09).
Furthermore, unlike BL Lacs, WLQs do not display significant polarization or variability (DS09).
These properties alone are a strong, but not conclusive, argument against the BL Lac hypothesis for explaining the weak lines in WLQs, given that radio-quiet BL Lac objects must be extraordinarily rare, if they exist at all (Stocke \et 1990, and see also Januzzi \et 1993, Londish \et 2004, and Plotkin \et 2010b).

The `BL Lac' and `extreme quasar' scenarios provide differing predictions for the shape of the UV-to-mid-infrared (IR) spectral energy distributions (SEDs) of WLQs.
The BL Lac emission in this range is mostly due to synchrotron emission; the synchrotron radiation peaks either in the UV/soft-\xray\ bands for high-energy peaked BL Lacs (HBLs), or in the near-IR band for low-energy peaked BL Lacs (LBLs; e.g., Padovani \& Giommi 1995; Fossati \et 1998; Nieppola \et 2006).
The prominent LBL near-IR peak dominates the UV-to-mid-IR SED, while an HBL-like SED in this spectral range is characterized by a power-law representing the tail of the synchrotron peak. In contrast, ordinary quasars' SEDs display the characteristic `dip'
at $\sim1~\mu$m between the emissions from the disk and the circumnuclear heated dust (e.g., Elvis \et 1994).
DS09 traced the UV-to-mid-IR SEDs of four WLQs (SDSS~J1302$+$0030 at $z=4.5$, SDSS~J1408$+$0205 at $z=4.0$,
SDSS~J1442$+$0110 at $z=4.5$, and SDSS~J1532$-$0039 at $z=4.6$)
using \spitzer\ Space Telescope mid-IR photometry as well as archival SDSS spectra and ground-based, near-IR photometry.
All their SEDs deviate significantly from a pure power-law, HBL-like continuum, and none can be fitted with an LBL-like spectrum.
Instead, their results indicate the presence of hot ($T\sim1000$\,K) dust emission in all four sources, consistent with the SEDs of
ordinary quasars.

In this work, we extend the near-to-mid-IR photometry to a statistically representative sample of 18 WLQs, including the four DS09 sources, and present conclusive evidence against the possibility that a relativistically-boosted continuum dilutes the WLQ line emission.
We describe our observations in \S~\ref{sec:observations} and present our basic findings, including a composite UV-to-mid-IR SED of WLQs in \S~\ref{sec:results}. We discuss our results in \S~\ref{sec:discussion}, and present our conclusions in \S~\ref{sec:summary}.
Throughout the text, shortened versions of object names are used and full names appear in the tables and figures. Luminosity distances are computed using the standard cosmological model (\hbox{$\Omega _{\Lambda}= 0.7$}, \hbox{$\Omega _{\rm M}=0.3$},
and \hbox{$H _{0}=70$\,km s$^{-1}$ Mpc$^{-1}$}).

%%%
%%%
%%%
\section{Observations and Data Reduction} \label{sec:observations}
%%%
\subsection{New \spitzer\ Observations} \label{spitzer_obsv}
\begin{figure*}
\epsscale{1.03}
\plotone{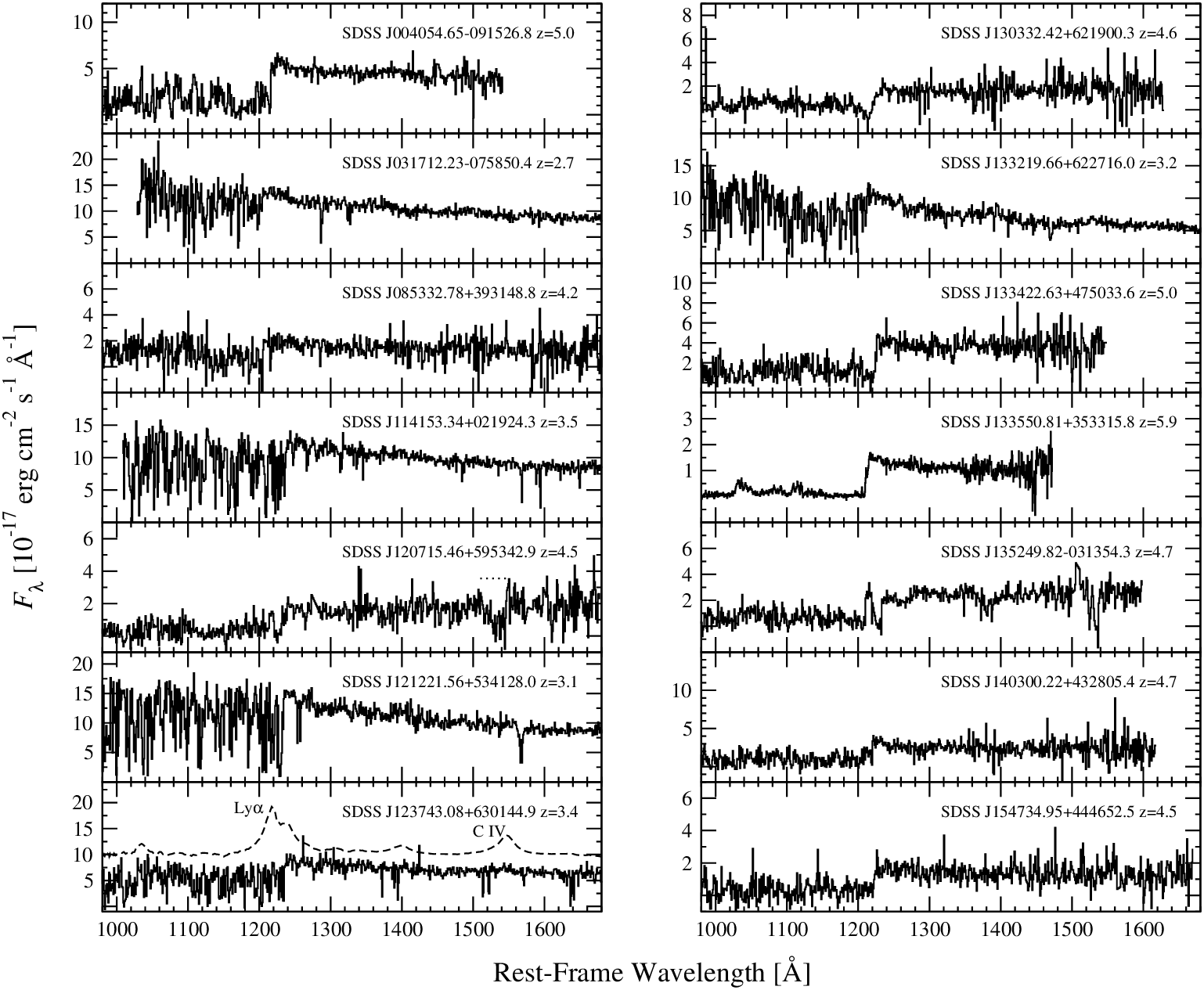}
\caption
{Rest-frame UV spectra of the WLQs selected for new \spitzer\ observations.
The spectra were obtained from the SDSS except for SDSS~J13350.81$+$353315.8, for which the spectrum was obtained from the Multiple Mirror Telescope.
The mean UV spectrum of $\sim2000$ SDSS quasars adapted from Vanden Berk \et (2001), scaled arbitrarily in flux, is shown
for comparison in the bottom left panel (dashed line) with prominent emission lines marked. The dotted line plotted on top of the
spectrum of SDSS~J120715.46$+$595342.9 marks the \ion{C}{4} BAL trough.
}
\label{sdss}
\end{figure*}

%%%
We have selected a representative sample of 14 high-redshift (\hbox{$2.7 \leq z \leq 5.9$}) WLQs that were bright enough for economical \spitzer\ Space Telescope (Werner \et 2004) photometry\footnote{Source selection was based on our early SDSS measurements; only seven of these sources are formally classified as WLQs by the DS09 study of SDSS quasars at $3.00\leq z\leq5.41$; see Table~\ref{target_props}.}.
Except for SDSS~J1335$+$3533, for which the rest-frame UV spectrum was obtained from Fan \et (2006), our sources' spectra were obtained from the SDSS Data Release~3 quasar catalog (Schneider \et 2005). These sources
represent the ranges in luminosity and redshift as well as the range in radio loudness observed in the WLQ population (none of our sources are radio loud, i.e., all have $R<100$, where
\hbox{$R=f_{\rm 5\,GHz}/f_{\rm 4400\,\mbox{\scriptsize\AA}}$}; Kellermann \et 1989). In fact, our sample constitutes the majority of the WLQ population that were known at the time; it also remains representative of the
larger WLQ population known today (e.g., DS09; Plotkin \et 2010a).
Basic properties of the sources are given in Table~\ref{target_props}, and their observed-frame UV spectra are displayed in
Fig.~\ref{sdss}.

%%%
%%%
%%%
%%%
\begin{deluxetable}{lcccccc}
\tabletypesize{\scriptsize}
\tablecaption{Basic Properties of the New WLQ Sample}
\tablewidth{0pt}
\tablehead{
\colhead{Quasar} &
\colhead{} &
\colhead{} &
\colhead{$\log \nu L_{\nu}$\tablenotemark{a}} &
\colhead{} &
\colhead{} \\
\colhead{(SDSS J)} &
\colhead{$z$} &
\colhead{$AB_{1445(1+z)\,\rm \scriptsize \AA}$} &
\colhead{(erg~s$^{-1}$)} &
\colhead{$R$\tablenotemark{b}} &
\colhead{Reference}
}
\startdata
004054.65$-$091526.8\tablenotemark{d}  & 5.0 & 18.79 & 47.0 & $<3$                                   & 1 \\
031712.23$-$075850.4                                  & 2.7 & 18.68 & 46.6 & $<12$\tablenotemark{c} & 1 \\
085332.78$+$393148.8                                 & 4.2 & 20.24 & 46.3 & 35                                        & 2 \\
114153.34$+$021924.3\tablenotemark{d} & 3.5 & 18.50 & 46.9 & 12                                        & 1 \\
120715.46$+$595342.9                                 & 4.5 & 19.95 & 46.5 & $<8$                                   & 2 \\
121221.56$+$534128.0\tablenotemark{d} & 3.1 & 18.63 & 46.7 & $<2$                                   & 1 \\
123743.08$+$630144.9\tablenotemark{d} & 3.4 & 18.97 & 46.7 & $<3$                                   & 1 \\
130332.42$+$621900.3\tablenotemark{d} & 4.6 & 20.10 & 46.4 & $<10$                                 & 2 \\
133219.66$+$622716.0                                 & 3.2 & 19.17 & 46.5 & 33                                        & 1 \\
133422.63$+$475033.6\tablenotemark{d} & 5.0 & 18.92 & 46.9 & $<4$                                   & 2 \\
133550.81$+$353315.8                                 & 5.9 & 19.98 & 46.6 & $<9$                                   & 3 \\
135249.82$-$031354.3                                  & 4.7 & 19.40 & 46.7 & $<5$                                   & 2 \\
140300.22$+$432805.4\tablenotemark{d} & 4.7 & 19.32 & 46.7 & $<5$                                   & 2 \\
154734.95$+$444652.5                                 & 4.5 & 19.82 & 46.5 & $<8$                                   & 2
\enddata
\tablenotetext{a}{Monochromatic luminosity averaged over a 40\,\AA-wide bin centered on a rest-frame wavelength of 1445\,\AA,
corrected for Galactic extinction.}
\tablenotetext{b}{Radio loudness parameter; see \S~\ref{spitzer_obsv}.
Flux densities at a rest-frame wavelength of 4400\,\AA\ were obtained by extrapolation of the flux densities at a rest-frame wavelength
of 1445\,\AA\ assuming an optical continuum of the form \hbox{$f_{\nu} \propto \nu^{-0.5}$}.
Unless otherwise noted, flux densities at a rest-frame frequency of 5~GHz were computed from the flux densities at an observed frame wavelength of 20~cm, obtained from the Faint Images of the Radio Sky at Twenty-Centimeters (FIRST) survey (Becker \et 1995), using
a radio continuum of the form $f_{\nu} \propto \nu^{-0.5}$; upper limits on $R$ were calculated according to the $3~\sigma$
FIRST detection threshold at the source position.}
\tablenotetext{c}{Upper limit was calculated according to the NRAO/VLA Sky Survey (Condon \et 1998) detection threshold of 2.5~mJy.}
\tablenotetext{d}{Source formally classified as a WLQ at $3.00\leq z\leq5.41$ by DS09 and is listed in their Table~2.}
\tablerefs{(1) Collinge \et 2005; (2) Schneider \et 2005; (3) Fan \et 2006.}
\label{target_props}
\end{deluxetable}

%%%
%%%
%%%

We conducted \spitzer\ observations of the 14 WLQs using the InfraRed Array Camera (IRAC; Fazio \et 2004) and the Multiband Imaging Photometer for \spitzer\ (MIPS; Rieke \et 2004) in Cycle~3 (Program ID \# 30476).  
Each WLQ was observed with IRAC in the 3.6~$\mu$m, 4.5~$\mu$m, 5.8~$\mu$m, and 8.0~$\mu$m bands and with MIPS in the
24~$\mu$m band.
Total integration times for the IRAC exposures range between $\sim400$ and $\sim2000$~s with 100~s per frame, per source; the total integration times for the MIPS exposures range between $\sim400$ and $\sim2100$~s with 30~s per frame, per source.
These exposures were designed to achieve a signal-to-noise ratio of $\sim20$ for each channel.
The \spitzer\ observation log is presented in Table~\ref{obsv_log}.

Our analysis of the IRAC and MIPS observations is based on the Basic Calibrated Data (BCD) images from version S18.5.0 of the
\spitzer\ Science Center (SSC) pipeline.
The \spitzer-specific software package \mopex\ (version 18.2.2; Makovoz \& Marleau 2005) was used to generate mosaics from the BCD images using background matching and outlier pixel rejection.
These mosaics excluded the first image for each IRAC channel and the first three images for MIPS, for each target, due to the `first-frame effect' (see, e.g., the IRAC Instrument Handbook\footnote{http://ssc.spitzer.caltech.edu/irac/iracinstrumenthandbook/home} and the MIPS Instrument Handbook\footnote{http://ssc.spitzer.caltech.edu/mips/mipsinstrumenthandbook/home}).
Aperture photometry was performed on the new mosaics using \apex, a \spitzer-specific  photometry routine incorporated into \mopex, using, for all IRAC channels, a source aperture radius of 3.6\arcsec\ (3 pixels) with the background measured in a source-free annulus (centered on the source) with an inner radius of 14.4\arcsec\ and a width of 7.2\arcsec; and using, for the MIPS 24 $\mu$m channel, a source aperture radius of 6.1\arcsec\ (2.5 pixels) with the background measured in a source-free annulus (centered on the source) with an inner radius of 20\arcsec\ and a width of 12.2\arcsec.

%%%
%%%
%%%
%%%
\begin{deluxetable*}{lccrrrrccc}
%
%\rotate
\tabletypesize{\scriptsize}
\tablecolumns{10}
\tablewidth{0pt}
\tablecaption{\spitzer\ Observation Log}
\tablehead{
\colhead{Quasar}	&
\colhead{}                   &
\colhead{}          	&
\multicolumn{4}{c}{IRAC Integration Time (s)} \\
\cline{4-7} &
\colhead{}		&
\colhead{}         &
\colhead{}		&
\colhead{}		&
\colhead{}		&
\colhead{}		&
\colhead{}          &
\colhead{}          &
\colhead{MIPS $24~\mu$m} \\
\colhead{(SDSS J)}	&
\colhead{IRAC Campaign}	&
\colhead{MJD }		&
\colhead{$3.6~\mu$m}	&
\colhead{$4.5~\mu$m}   & 
\colhead{$5.8~\mu$m}	&
\colhead{$8.0~\mu$m}	&
\colhead{MIPS Campaign}	&
\colhead{MJD}          	&
\colhead{Integration Time (s)}
}
\startdata
004054.65$-$091526.8		&		IRAC008900		&		53960.02		&	774		&	871		&		774		&		796  	&		MIPS007600	&		53939.36		&	742		\\
031712.23$-$075850.4		&		IRAC009500		&		54147.18		&	484		&	581		&		484		&		515		&		MIPS012600	&		54339.14		&	402		\\
085332.78$+$393148.8		&		IRAC009700		&		54227.04		& 1452 	& 	1531	&		1452	&		1404	&		MIPS010700	&		54076.87		&	1980 	\\
114153.34$+$021924.3		&		IRAC009800		&		54284.85		&	387		&	484		&		387		&		421		&		MIPS012000	&		54292.42		&	371		\\
120715.46$+$595342.9		&		IRAC009200		&		54069.68		&	1452	&	1549	&		1403	&		1451	&		MIPS010700	&		54077.76		&	2103  	\\
121221.56$+$534128.0		&		IRAC009300		&		54096.08		&	387		&	484		&		387		&		421		&		MIPS010700	&		54077.71		&	402		\\
123743.08$+$630144.9		&		IRAC009200		&		54066.94		&	484		&	580		&		484		&		515		&		MIPS010700	&		54077.70		&	402		\\
130332.42$+$621900.3		&		IRAC009200		&		54069.65		&	1161	&	1259	&		1161	&		1170	&		MIPS010700	&		54077.68		&	1701	\\
133219.66$+$622716.0		&		IRAC009300		&		54095.59		&	580		&	678		&		581		&		608		&		MIPS010700	&		54077.67		&	804		\\
133422.63$+$475033.6		&		IRAC009500		&		54150.81		&	968		&	1065	&		968		&		983		&		MIPS011900	&		54249.74		&	1238	\\
133550.81$+$353315.8		&		IRAC009500		&		54151.25		&	1937	&	2032	&		1936	&		1918	&		MIPS012000	&		54305.04		&	2567	\\
135249.82$-$031354.3		&		IRAC008900		&		53961.30		&	968		&	1065	&		968		&		983		&		MIPS007600	&		53938.85		&	1237	\\
140300.22$+$432805.4		&		IRAC009500		&		54150.23		&	1161	&	1259	&		1161	&		1170	&		MIPS011900	&		54250.75		&	1640	\\
154734.95$+$444652.5		&		IRAC008900		&		53961.23		&	1161	&	1258	&		1161	&		1170	&		MIPS011800	&		54203.29		&	1670	
\enddata
\label{obsv_log}
\end{deluxetable*}
 
The IRAC fluxes were corrected for source aperture size using the aperture correction factors from the IRAC Instrument Handbook.
Array-location dependent corrections were  applied to the IRAC data but no color or pixel-phase corrections were applied as these were found to have an effect of $<1$\% on the total flux. For the MIPS data, aperture corrections were calculated using the MIPS 24~$\mu$m point spread function (PSF), and the fluxes were further increased by 4\% to correct for an $f_{\nu} \propto \nu^{-1}$ power law which represents an ordinary quasar SED in the $\sim4-6~\mu$m rest-frame band (corresponding to 24~$\mu$m in the observed frame for the redshifts of our sources; e.g., Elvis \et 1994; Richards \et 2006, hereafter R06). We note, however, that our main results, presented in \S~\ref{sec:results}, would not have changed significantly had this small correction not been applied. Table \ref{all_quasars} presents the results of our \spitzer\ photometry. The statistical flux uncertainties for all mosaics are smaller than the 5\% calibration accuracy of IRAC and the 4\% calibration accuracy of MIPS; however, we quote a conservative 5\% uncertainty for all our flux density values in Table~\ref{all_quasars}.

%%%
%%%
%%%
\subsection{Archival \spitzer\ Data} \label{sec:archivedata}
%%%
We also include \spitzer\ photometry of the four WLQs from DS09 in Table~\ref{all_quasars}.  We have reanalyzed the \spitzer\ data for three of these sources using the processes described in \S~\ref{spitzer_obsv}.
A nearby star prevents accurate aperture photometry on SDSS~J1532$-$0039 (see DS09); thus we performed PSF fitting for this source employing standard {\sc iraf}\footnote{Image Reduction and Analysis Facility (IRAF) is distributed by the
National Optical Astronomy Observatory, which is operated by AURA, Inc.,
under cooperative agreement with the National Science Foundation.} photometry tasks on the mosaics generated by \mopex.
The only significant difference we find with the DS09 fluxes for their four WLQs is for the MIPS 24~$\mu$m flux of the faintest source,
SDSS~J1442$+$0110, which we find to be higher by a factor of 1.7 in our new analysis.
We attribute this difference to an improved mosaicking technique, which uses an updated version of the SSC pipeline;
this discrepancy does not affect the main conclusions of either study.

%%%
%%%
%%%
\subsection{New Near-IR Observations} \label{sec:new_nir}
%%%
Near-IR observations were taken with the Astrophysical Research Consortium 3.5~m telescope at the Apache Point Observatory for two of our sources, SDSS J1212+5341 (2008 June 15 and 2009 February 14) and SDSS J1237+6301 (2009 February 14).
Images were taken in the $J$, $H$, and $K_{\rm S}$ filters with the Near-Infrared Camera \& Fabry-Perot Spectrometer.
In the $J$ filter, $15\times60$~s sub-exposures were taken over a five point dither pattern, for a total exposure time of 900~s for each source on each night. For the $H$ and $K_{\rm S}$ filters, 12~s sub-exposures were used over two sets of dither patterns with 25 sub-exposures each for a total exposure time of 600~s per filter on each night. Each sub-exposure was Fowler sampled eight times to reduce the read noise. Sky maps were created for each dither pattern by taking the median of the sub-exposures. Both 12~s and 60~s dark images were taken at the beginning of each night. To create the flat field, we subtracted the dark images from the sky maps, and then normalized them so the flat field had an average value of 1 count. After sky subtraction and flat fielding, the images were aligned (using 3--4 reference stars) and then co-added with standard tasks in {\sc iraf}. Differential photometry was then performed, using the Two Micron All Sky Survey (Skrutskie \et 2006) photometry of four field stars in the $4.6' \times 4.6'$ field-of-view of each source. Each field star was chosen to be near the target WLQ and of similar brightness. Finally, we corrected the photometric data for Galactic extinction by applying the standard extinction curve from Cardelli \et (1989), using the extinction maps of Schlegel \et (1998).
The $J$, $H$, and $K_{\rm S}$ photometric data for these two WLQs are presented in Table~\ref{nirdata}. 

%%%
%%%
%%%
%%%
\begin{deluxetable*}{lccccc}
\tabletypesize{\scriptsize}
\tablecaption{\spitzer\ Photometry for the Entire WLQ Sample}
\tablewidth{0pt}
\tablehead{ 
\colhead{Quasar} &
\multicolumn{5}{c}{Flux Density ($\mu$Jy)} \\
\cline{2-6} &
\colhead{} &
\colhead{} &
\colhead{} &
\colhead{} &
\colhead{} \\
\colhead{(SDSS J)} &
\colhead{$3.6~\mu$m} &
\colhead{$4.5~\mu$m} &
\colhead{$5.8~\mu$m} &
\colhead{$8.0~\mu$m} &
\colhead{$24.0~\mu$m} 
}
\startdata
004054.65$-$091526.8	&    \phn 101.0$\pm$5.1	&	\phn 79.6	$\pm$4.0	&	\phn 75.0	$\pm$3.8	&	\phn 110.0$\pm$5.5	&	\phn 509$\pm$	26 \\
031712.23$-$075850.4	&    \phn 135.0$\pm$6.8	&	\phn 141.0$\pm$7.1	&	\phn 186.0$\pm$9.3	&	\phn 335$\pm$17	&	\phn 610$\pm$	31 \\
085332.78$+$393148.8	&    \phn 38.0$\pm$1.9	&	\phn 29.0$\pm$1.5	&	\phn 36.6$\pm$1.8	&	\phn 53.5$\pm$2.7	&	\phn 605$\pm$	30 \\
114153.34$+$021924.3	&    \phn 236$\pm$12	&	\phn 235$\pm$12	&	\phn 295$\pm$15	&	\phn 499$\pm$25	&	\phn 1380	$\pm$140 \\
120715.46$+$595342.9	&    \phn 167.0$\pm$8.3	&	\phn 146.0$\pm$7.3	&	\phn 174.0$\pm$8.7	&	\phn 324$\pm$16	&	\phn 1500	$\pm$75 \\
121221.56$+$534128.0	&    \phn 239$\pm$12	&	\phn 281$\pm$14	&	\phn 405$\pm$20	&	\phn 835$\pm$42	&	\phn 2650	$\pm$130 \\
123743.08$+$630144.9	&    \phn 132.0$\pm$6.6	&	\phn 139$\pm$7	&	\phn 197.0$\pm$9.9	&	\phn 405$\pm$20	&	\phn 1450	$\pm$73 \\
130216.13$+$003032.1\tablenotemark{a}	&     \phn 75.2$\pm$3.8	&	\phn 58.1$\pm$2.9	&	\phn 60.2$\pm$3.0	&	\phn \phn 79.1$\pm$4.0	&	\phn 420$\pm$21 \\
130332.42$+$621900.3	&    \phn 89.2$\pm$4.5	&	\phn 72.2$\pm$3.6	&	\phn 95.1$\pm$4.8	&	\phn 151.0$\pm$7.6	&	\phn 552$\pm$28 \\
133219.66$+$622716.0	&    \phn 103.0$\pm$5.1	&	\phn 102.0$\pm$5.1	&	\phn 127.0$\pm$6.4	&	\phn 212$\pm$11	&	\phn 753$\pm$38 \\
133422.63$+$475033.6	&    \phn 120$\pm$6 	     &	\phn 97.7$\pm$4.9	&	\phn 107.0$\pm$5.3	&	\phn 152.0$\pm$7.6	&
\phn 901$\pm$45 \\
133550.81$+$353315.8	&    \phn 67.1$\pm$3.4	&	\phn 67.8$\pm$3.4	&	\phn 50.5$\pm$2.5	&	\phn 54.8$\pm$2.8	&	\phn 417$\pm$21 \\
135249.82$-$031354.3	&    \phn 165.0$\pm$8.2	&	\phn 151.0$\pm$7.5	&	\phn 170.0$\pm$8.5	&	\phn 261$\pm$13	&	\phn 1370	$\pm$68 \\
140300.22$+$432805.4	&    \phn 90.7$\pm$4.5	&	\phn 67.8$\pm$3.4	&	\phn 82.3$\pm$4.1	&	\phn 136.0$\pm$6.8	&	\phn 796$\pm$40 \\
140850.92$+$020522.7\tablenotemark{a}	&     \phn 87.1$\pm$4.4	&	\phn 73.1$\pm$3.7	&	\phn 73.0$\pm$3.7	& \phn 141.0$\pm$7.1	&	\phn 530$\pm$27 \\
144231.72$+$011055.3\tablenotemark{a}	&     \phn 37.8$\pm$1.9	&	\phn 39.8$\pm$2.0	&	\phn 34.1$\pm$1.7	&	\phn 69.4$\pm$3.5	&	\phn 325$\pm$16 \\
153259.96$-$003944.1\tablenotemark{a,b}	&     \phn 81.0$\pm$4.1	&	\phn 72.6$\pm$3.6	&	\phn 83.2$\pm$6.1	&	\phn 123.0$\pm$6.1	&	\phn 502$\pm$25 \\
154734.95$+$444652.5	&     \phn 84.6$\pm$4.2	&	\phn 71.3$\pm$3.6	&	\phn 72.3$\pm$3.6	&	\phn 106.0$\pm$5.3	&	\phn 406$\pm$20
\enddata
\tablenotetext{a}{WLQ from DS09.}
\tablenotetext{b}{Flux densities from PSF photometry; see \S~\ref{sec:archivedata} for details.}
\label{all_quasars}
\end{deluxetable*}

%%%
%%%
%%%
\section{Spectral Energy Distribution of WLQs} \label{sec:results}
%%%
\subsection{UV-to-Mid-IR SEDs of Individual WLQs} \label{sec:seds}
\begin{figure*}
\epsscale{1.0}
\plotone{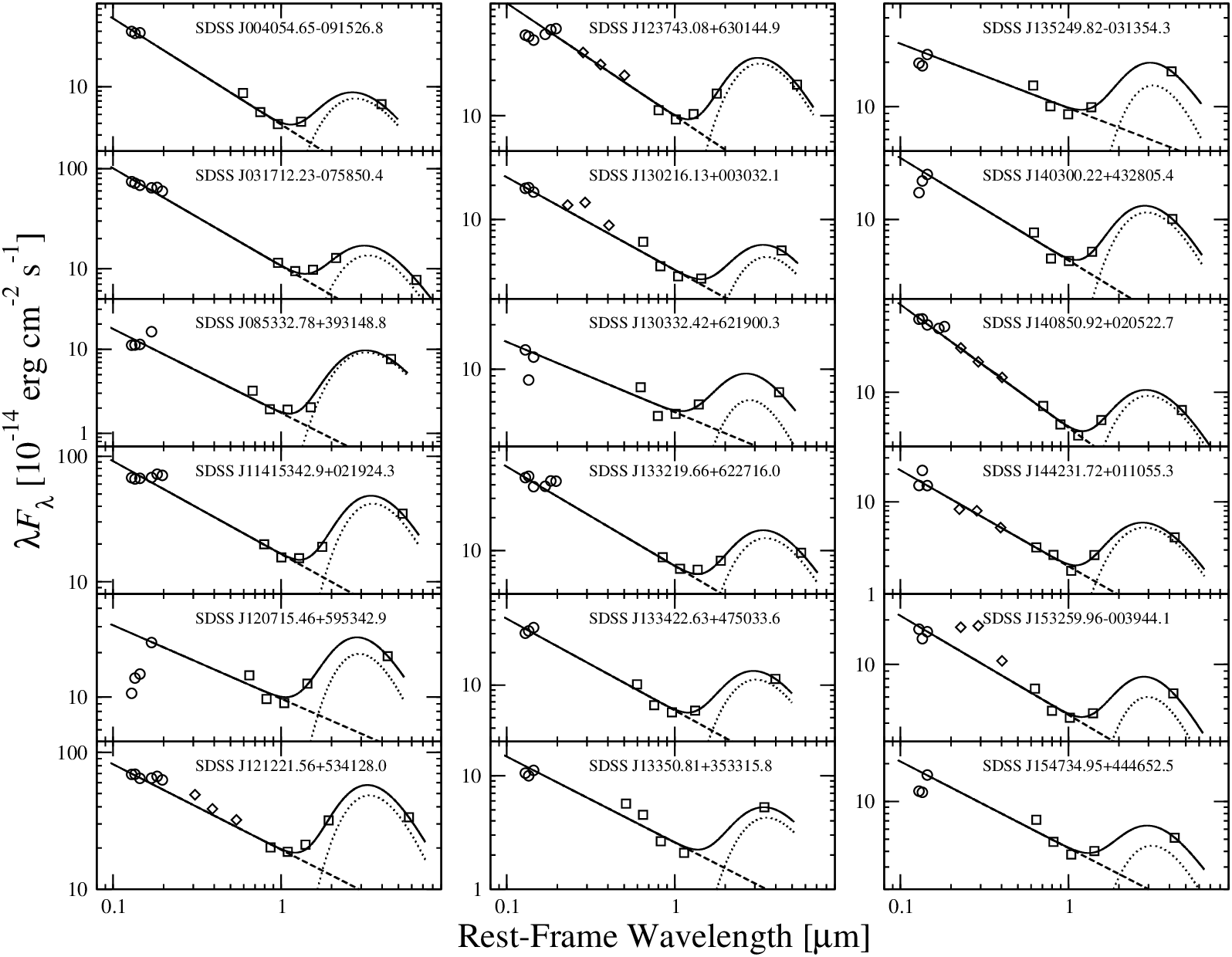}
\caption{Rest-frame UV-to-mid-IR SEDs of our WLQs.
The rest-frame UV bins are marked with circles, the observed-frame near-IR
photometry, where available, are marked with diamonds, and the \spitzer\ photometric data are marked with squares; symbol sizes are larger than the uncertainties on the data. A best-fit power-law model (dashed line), a blackbody model (dotted line), and a power law plus
blackbody model (solid line) for each SED is shown in each panel; the best-fit parameters are listed in Table~\ref{stats}.
Note the reddened SED of the BAL quasar SDSS~J120715.46$+$595342.9.}
\label{sed_all}
\end{figure*}

%%%
In Fig.~\ref{sed_all} we present the UV-to-mid-IR SEDs of the 18 WLQs observed by \spitzer. The SEDs are composed of the
\spitzer\ photometry from Table~\ref{all_quasars} as well as near-IR photometry for six of the sources; two sources from Table~\ref{nirdata} and four sources from DS09\footnote{The near-IR photometry for the four DS09 sources was corrected for Galactic extinction following the procedures outlined in \S~\ref{sec:new_nir}.} (see \S~\ref{sec:introduction}). The rest-frame UV spectral band in the SEDs was obtained from the WLQs' SDSS spectra, with the exceptions of SDSS~J1532$-$0039 and SDSS~J1335$+$3533, for which the UV bands were obtained from spectra taken with the Low-Resolution Imaging Spectrometer at the W.~M.~Keck Observatory (Oke \et 1995) and with the Red Channel Spectrograph\footnote{http://www.mmto.org/instruments/rcupdate.shtml} on the Multiple Mirror Telescope, respectively.
The rest-frame UV spectra were corrected for Galactic extinction using the same procedures as in \S~\ref{sec:new_nir}.
The flux densities were averaged over line-free, 40\,\AA-wide bands centered at rest-frame wavelengths 1290\,\AA, 1350\,\AA, 1445\,\AA, 1695\,\AA, 1830\,\AA, and 1965\,\AA, depending on source redshift (see Table~\ref{sdss_binned}).
Table~\ref{sdss_binned} also provides the power-law indices ($\alpha_{\nu}$) calculated from fitting a power law of the form
$f_{\nu} \propto \nu^{-\alpha_{\nu}}$ to the average fluxes in the rest-frame UV photometric bins.
Our $\alpha_{\nu}$ values lie in the range \hbox{$0.28<\alpha_{\nu}<2.90$}
with a mean and standard deviation of 1.12 and 0.73, respectively. These WLQs therefore have
rest-frame UV continua that are consistent, on average,
with the respective continua of ordinary SDSS quasars at \hbox{$3.6 < z < 5.0$} (comparable to the redshift range of our WLQs)
that show power-law indices of $\alpha_{\nu}=0.79\pm0.34$ (Fan \et 2001).

\begin{figure*}
\epsscale{1.0}
\plotone{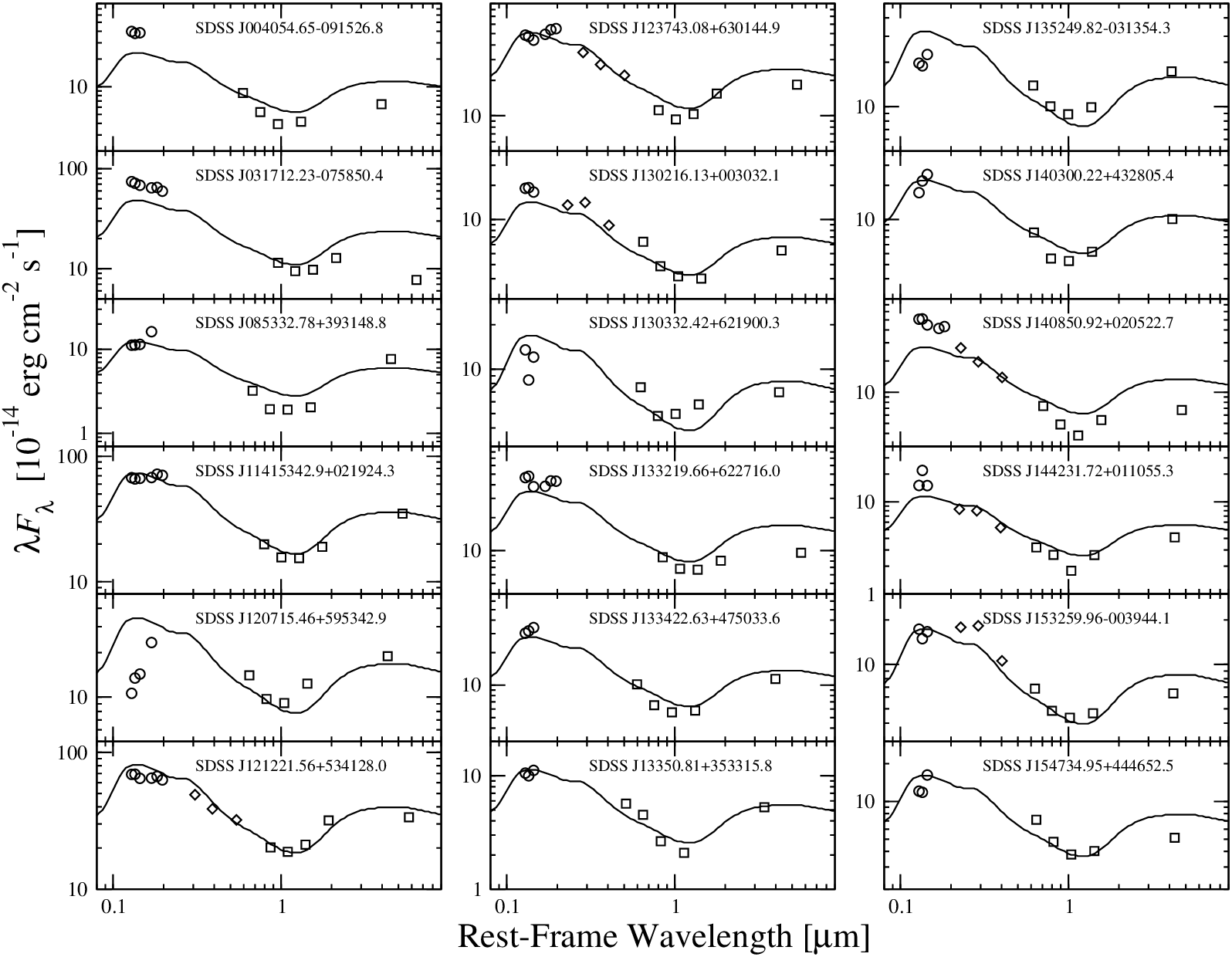}
\caption{Rest-frame UV-to-mid-IR SEDs of our WLQs. The linearly-scaled composite SED of optically-luminous quasars from R06 (solid line) is shown for comparison in each panel (symbols are the same as in Fig.~\ref{sed_all}). Note the discrepancy between the R06 composite SED and the SED of the BAL quasar SDSS~J120715.46$+$595342.9 in the rest-frame UV band.}
\label{fit_seds}
\end{figure*}

Inspection of Fig.~\ref{sed_all} clearly indicates that a power-law model\footnote{Where the power-law index, $\alpha_{\lambda}$, is defined as $\lambda f_{\lambda} \propto \lambda ^{-\alpha_{\lambda}}$; see Table~\ref{stats}.} cannot represent the entire UV-to-mid-IR SED of any of our WLQs; the fluxes at the longest wavelengths are significantly above the power-law SED in each source.
To account for this deviation, we added a single blackbody component to the power-law model; the best-fit models are shown on top of the WLQ SEDs in Fig.~\ref{sed_all} and Table~\ref{stats} presents the best-fit parameters.
With the exception of SDSS~J1207$+$5953,
the best-fit power-law indices are in the range \hbox{$0.42\leq \alpha_{\lambda} \leq 1.21$} and the best-fit blackbody temperatures are in the range \hbox{$820\leq T\leq1050$\,K}.
Due to the limited number of photometric data points available for the fits, the best-fit parameters have an uncertainty of $\sim 15$\%. 
Fig.~\ref{fit_seds} shows a composite SED of optically-luminous quasars from R06 plotted on top of the SED of each WLQ in our sample;
the R06 composite SED was vertically shifted to provide the best-fit model to each WLQ SED. It is apparent that every WLQ SED is, qualitatively, consistent with the SED of an ordinary quasar.
SDSS~J1207$+$5953 with \hbox{$\alpha_{\lambda} = 0.04$} may suffer from intrinsic reddening,
and the weakness of its emission lines may be at least partly attributed to this effect (see also Figs.~\ref{sed_all}~and~\ref{fit_seds}). 
Furthermore, this source has recently been found to show a hint of a broad \ion{C}{4} absorption trough and was therefore classified as
a BAL quasar\footnote{The possible BAL quasar nature of this source was not known during the \spitzer\ target selection process.} (Trump \et 2006; Gibson \et 2009; see Fig.~\ref{sdss}); this is the only known BAL quasar in our sample.
We have therefore excluded this source from our composite WLQ SED (\S~\ref{sec:mean_sed}).

We note that the rest-frame UV spectra, near-IR photometry, and \spitzer\ photometry were obtained at different epochs, spaced by
$\approx 1$~yr in the rest frame.
These temporal gaps, however, should not affect the shapes of individual SEDs significantly assuming that WLQs vary as 
luminous, high-redshift quasars (which do not vary by more than $\sim10$\% on such timescales; e.g., Kaspi \et 2007; see also
\S~\ref{sec:introduction} and DS09) and considering that the uncertainties on our fluxes are on the order of $\sim5$\%.

%%%
%%%
%%%
%%%
\begin{deluxetable}{lcccc}
\tabletypesize{\scriptsize}
\tablecaption{New Near-IR Photometry from the ARC 3.5\,m Telescope}
\tablewidth{0pt}
\tablehead{ 
\colhead{Quasar} &
\colhead{$J$} &
\colhead{$H$} &
\colhead{$K_{\rm S}$}\\
\colhead{(SDSS J)} &
\colhead{(mag)} &
\colhead{(mag)} &
\colhead{(mag)}
}
\startdata
121221.56$+$534128.0	&	17.23	$\pm$ 0.02	&   16.80	$\pm$ 0.03	&   16.13	$\pm$ 0.08\\
123743.08$+$630144.9	&	17.61	$\pm$ 0.02	&   17.18	$\pm$ 0.04	&   16.53	$\pm$ 0.09
\enddata
\label{nirdata}
\end{deluxetable}

%%%
%%%
%%%
\subsection{Composite WLQ SED} \label{sec:mean_sed}
%%%
We create a composite WLQ UV-to-mid-IR SED from the individual SEDs of all our sources, excluding SDSS~J1207$+$5953
(see \S~\ref{sec:seds}). The data from the 17 WLQs were binned into six rest-frame UV, two near-IR (i.e., the ground-based observations), and seven near-to-mid-IR (i.e., \spitzer) flux bins.
The seven \spitzer\ bins include five covering the four IRAC bands, with a minimum of 10 sources per bin, and two bins covering the MIPS~24\,$\mu$m band with a minimum of 8 sources per bin.
The flux of the composite SED in each bin represents the flux-weighted mean of all the individual SEDs contributing to that bin
using the average flux densities in the 40\,\AA-wide bin centered on rest-frame 1445\,\AA\ in each source.
The fluxes of the composite SED were normalized by the mean flux density at rest-frame 1445\,\AA.
The central wavelength of each \spitzer\ bin is the flux-density weighted mean
rest-frame wavelength of the data points contributing to that bin.
The composite WLQ SED data are provided in Table~\ref{mean_sed_table} and shown in Fig.~\ref{mean_sed}.

Fig.~\ref{mean_sed} shows that the composite WLQ SED is well represented by a model consisting of a single power law and a single blackbody; a least-squares fit results in a power-law index of $\alpha_{\lambda}=1.09\pm0.16$ and a blackbody temperature of
$T=960 \pm140$\,K; the separate best-fit model components, i.e., a power law and a blackbody, are also shown.
Also shown in Fig.~\ref{mean_sed} are the R06 composite SED of optically-luminous quasars and a parabolic model representing
a typical LBL SED from Nieppola \et (2006); it is clear that the WLQ SED cannot be represented by an LBL SED. The R06
quasar SED provides a good representation of the composite WLQ SED at all wavelengths, deviating in flux by only $\sim$10\% in the
$J$ and MIPS~24~$\mu$m bands.

To obtain a comparison between the SEDs of WLQs and ordinary quasars, we created a
composite quasar SED using a subset of 22 R06 quasars that closely resemble the luminosity and redshift distributions of our WLQ sample. The redshift and luminosity ranges of this subset are \hbox{$2.13 \leq z \leq 5.22$} and
\hbox{$45.76 \leq \log \nu L _{\nu} (1445\,\mbox{\AA}) \leq 47.64$}, respectively.
The basic properties of the R06 quasar subset are given in Table \ref{rich_subset_props}.
The composite SED for this subset was created following the method described above for the composite WLQ SED.
The R06 quasar subset composite SED is shown with the composite WLQ SED in
Fig.~\ref{mean_sed}c.
Another comparison composite UV-to-mid-IR SED was created (as described above for the composite WLQ SED)
using 12 of the 13 $z \sim 6$ quasars from Jiang \et (2006).
SDSS~J000552.34$-$000655.8 was excluded from that sample because it does not have a MIPS~24~$\mu$m detection
(see also Jiang \et 2010). The Jiang \et (2006) composite SED is shown in Fig.~\ref{mean_sed}d.
The composite SEDs of both the R06 quasar subset and the Jiang \et (2006) sample are consistent, within the errors
(i.e., within $\sim5$\%), with our composite WLQ SED.

\begin{deluxetable*}{lccccccc}
\tabletypesize{\scriptsize}
\tablecaption{Rest-Frame UV Spectroscopy for the Entire WLQ Sample}
\tablewidth{0pt}
\tablehead{ 
\colhead{Quasar} &
\multicolumn{6}{c}{Flux Density ($\mu$Jy)} \\
\cline{2-7} &
\colhead{} &
\colhead{} &
\colhead{} &
\colhead{} &
\colhead{} &
\colhead{} &
\colhead{} \\
\colhead{(SDSS J)} &
\colhead{1290\,\AA} &
\colhead{1350\,\AA} &
\colhead{1445\,\AA} &
\colhead{1695\,\AA} &
\colhead{1830\,\AA} &
\colhead{1965\,\AA}	&
\colhead{$\alpha_{\nu}$\tablenotemark{a} }	
}
\startdata
004054.65$-$091526.8 &  102.3 $\pm$ 6.1 & 102.3 $\pm$ 9.3 & 111.0 $\pm$ 17.3 & \nodata &  \nodata   & \nodata   & 0.76 \\
031712.23$-$075850.4 & 118.3 $\pm$ 6.8 & 119.2 $\pm$ 9.5 & 121.9 $\pm$ 8.9 & 134.6 $\pm$ 8.4 & 146.6 $\pm$ 6.6 & 144.4 $\pm$ 14.8 & 0.55 \\
085332.78$+$393148.8 &  25.2 $\pm$ 10.3 & 26.5 $\pm$ 7.0 & 28.9 $\pm$ 12.9 & 47.9 $\pm$ 23.4 & \nodata   & \nodata   & 1.25 \\
114153.34$+$021924.3 &  129.9 $\pm$ 9.9 & 132.6 $\pm$ 19.8 & 143.9 $\pm$ 7.7 & 171.4 $\pm$ 6.8 & 196.6 $\pm$ 12.5 & 206.6 $\pm$ 15.2 & 1.16 \\
120715.46$+$595342.9 & 25.1 $\pm$ 3.3 & 33.2 $\pm$ 10.7 & 37.9 $\pm$ 15.6 & 73.4 $\pm$ 3.9 & \nodata   & \nodata   & 1.96 \\
121221.56$+$534128.0 & 121.6 $\pm$ 5.4 & 127.9 $\pm$ 12.0 & 127.5 $\pm$ 11.3 & 150.1 $\pm$ 11.6 & 166.9 $\pm$ 8.4 & 168.9 $\pm$ 18.0 & 0.83 \\
123743.08$+$630144.9 & 92.3 $\pm$ 10.2 & 94.1 $\pm$ 8.9 & 93.8 $\pm$ 6.3 & 123.1 $\pm$ 15.2 & 145.6 $\pm$ 16.2 & 159.6 $\pm$ 31.3 & 1.08 \\
130216.13$+$003032.1 &  44.2 $\pm$ 5.3 & 46.9 $\pm$ 4.8 & 45.9 $\pm$ 9.8 & 43.1 $\pm$ 12.1 & \nodata   & \nodata   & 0.28 \\
130332.42$+$621900.3 &  32.9 $\pm$ 8.0 & 21.6 $\pm$ 8.4 & 33.0 $\pm$ 11.6 & \nodata   & \nodata   & \nodata   & 0.51 \\
133219.66$+$622716.0 &  83.7 $\pm$ 6.5 & 90.2 $\pm$ 28.6 & 77.5 $\pm$ 9.6 & 91.6 $\pm$ 4.6 & 110.9 $\pm$ 8.3 & 118.3 $\pm$ 11.7 & 0.81 \\
133422.63$+$475033.6 &  77.5 $\pm$ 7.9 & 85.2 $\pm$ 14.2 & 97.9 $\pm$ 20.9 & \nodata   & \nodata   & \nodata   & 2.07 \\
133550.81$+$353315.8\tablenotemark{b} &  31.5 $\pm$ 4.3 & 31.1 $\pm$ 7.0 & 37.2 $\pm$ 15.5 & \nodata   & \nodata   & \nodata   & 1.48 \\
135249.82$-$031354.3 &  48.8 $\pm$ 7.1 & 49.1 $\pm$ 19.7 & 62.5 $\pm$ 9.9 & \nodata   & \nodata   & \nodata   & 2.19 \\
140300.22$+$432805.4 &  42.2 $\pm$ 14.1 & 56.0 $\pm$ 20.9 & 68.3 $\pm$ 31.1 & \nodata   & \nodata   & \nodata   & 2.90 \\
140850.92$+$020522.7 & 109.3 $\pm$ 11.3 & 115.2 $\pm$ 9.3 & 107.9 $\pm$ 10.7 & 117.0 $\pm$ 22.5 & 131.7 $\pm$ 48.2 & \nodata   & 0.43 \\
144231.72$+$011055.3 & 35.5 $\pm$ 5.6 & 54.4 $\pm$ 18.0 & 39.9 $\pm$ 12.0 & \nodata   & \nodata   & \nodata   & 0.57 \\
153259.96$-$003944.1\tablenotemark{c} & 41.9 $\pm$ 2.0 & 37.8 $\pm$ 9.3 & 45.1 $\pm$ 2.1 & \nodata   & \nodata   & \nodata   & 0.80 \\
154734.95$+$444652.5 & 28.7 $\pm$ 6.0 & 29.4 $\pm$ 8.1 & 43.2 $\pm$ 21.5 & \nodata   & \nodata   & \nodata   & 0.54 
\enddata
\tablecomments{Unless otherwise noted, the flux densities were obtained from SDSS spectra that were averaged over
\hbox{40\,\AA-wide} bins centered on the given wavelength; the uncertainties represent the standard deviation within the bin.}
\tablenotetext{a}{Power-law index from the best-fit function $f_{\nu} \propto \nu^{-\alpha_{\nu}}$ fitted to the flux densities.}  
\tablenotetext{b}{Spectrum obtained from the Multiple Mirror Telescope.}
\tablenotetext{c}{Spectrum obtained from the W.~M.~Keck Observatory.}
\label{sdss_binned}
\end{deluxetable*}

%%%
%%%
%%%
\section{Discussion} \label{sec:discussion}
%%%
Our \spitzer\ observations have increased the number of WLQ UV-to-mid-IR SEDs by more than a factor of four, allowing us to constrain their emission mechanism in this spectral range in a statistically meaningful manner.
By examining the WLQ SED in this critical spectral region we can determine if it is consistent with a relativistically-boosted continuum, e.g., that of a BL Lac object, or whether it shows signatures of a heated dust mid-IR `bump' characteristic of ordinary quasars.
Our WLQs have UV-to-mid-IR SEDs that are well fit by a model consisting of a power law plus blackbody with a power-law index of
$\alpha_{\lambda}\sim1$ and a blackbody temperature of $T \sim1000$\,K;
these parameters are typical of the emissions from a quasar accretion disk and circumnuclear heated dust, respectively (e.g., Elvis \et 1994; R06). None of the SEDs is consistent with pure synchrotron emission either from an HBL or an LBL, as all SEDs
show a significant departure from a pure power-law model at $\lambda_{\rm rest} \gtsim 1~\mu$m and none can be fitted with a parabolic function. These results set an upper limit of $\approx5$\% on the potential BL~Lac `contamination' fraction in WLQ samples.

\begin{figure*}
\epsscale{1.1}
\plotone{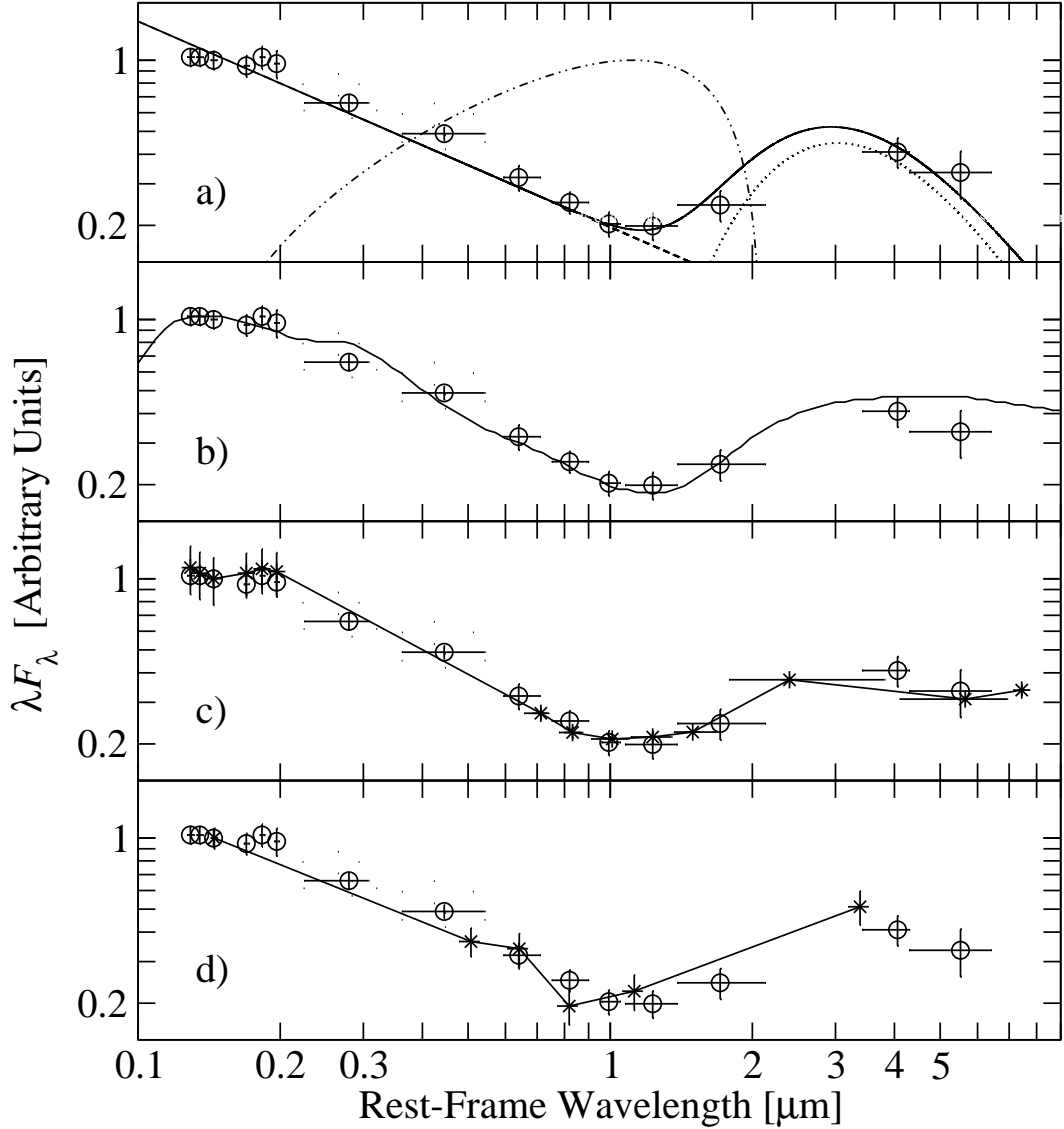}
\caption{Our composite WLQ SED (circles) is shown against ({\it a}) the best-fit model (solid line) composed of a power-law component(dashed line) and a single-blackbody component (dotted line) as well as an LBL SED model (shown for comparison; dot-dashed line), ({\it b}) the composite SED of optically-luminous quasars from R06 (solid line), ({\it c}) a composite SED assembled from a redshift and luminosity matched subset of R06 quasars (asterisks connected by a solid line), and ({\it d}) a composite SED assembled from the $z\sim6$ quasar sample of Jiang \et (2006, asterisks connected by a solid line). In each panel, the fluxes of all SEDs are normalized to the flux at 1445\,\AA.}
\label{mean_sed}
\end{figure*}

The R06 optically-luminous quasar composite SED provides an adequate fit to the composite WLQ SED, and
SED subsets composed from ordinary quasars with luminosities and redshifts similar to our WLQs (from R06 and Jiang \et 2006)  
are even better, matching our composite WLQ SED within the uncertainties (Fig.~\ref{mean_sed}).
This match provides conclusive evidence against the possibility that the emission lines in WLQs are overwhelmed by a relativistically-boosted continuum. The emission lines in WLQs must therefore be intrinsically weak.

%%%
%%%
%%%

%%%
%%%
%%%

The evidence from their UV-to-mid-IR SED, combined with their
\xray\ and radio properties (DS09; S09) as well as their variability and polarization properties (DS09),
demonstrates, unambiguously, that WLQs are unbeamed quasars with extreme emission line properties. 
The composite WLQ SED indicates that, at least in the rest-frame \hbox{$\sim0.1-5~\mu$m} spectral range, the broad-band properties of WLQs are indistinguishable from those of quasars with similar luminosities and redshifts, in spite of the significant difference in
emission-line strength.
While the reason for the discrepancy between the broad-band spectral properties and the emission-line properties in WLQs is still not understood, our results clearly show that WLQs cannot be selected based on their broad-band continuum properties alone, for example, by traditional photometric surveys in any band.
The selection of WLQs is currently feasible only with spectroscopic surveys such as the SDSS (e.g., Plotkin \et 2010a).

%%%
%%%
%%%
\begin{deluxetable}{lcccc}
\tablecaption{SED Best-Fit Parameters}
\tablewidth{0pt}
\tabletypesize{\scriptsize}
\tablehead{
\colhead{Quasar} &
\colhead{} &
\colhead{} &
\colhead{} &
\colhead{} \\
\colhead{(SDSS J)} &
\colhead{$\alpha_{\lambda}$\tablenotemark{a}} &
\colhead{$A_{\rm PL}$\tablenotemark{b}} &
\colhead{$T$\tablenotemark{c}} &
\colhead{$A_{\rm BB}$\tablenotemark{d}}
}
\startdata
004054.65$-$091526.8  & $1.17$ & $4.05$   & $1050$  & $7.50$   \\
031712.23$-$075850.4  & $0.96$ & $11.80$ & $880$    & $12.51$ \\
085332.78$+$393148.8 & $0.89$ & $1.92$   & $900$   &  $8.93$   \\
114153.34$+$021924.3 & $0.77$ & $17.69$ & $820$   &  $41.21$ \\
120715.46$+$595342.9 & $0.04$ & $11.21$ & $990$   &  $11.09$ \\
121221.56$+$534128.0 & $0.66$ & $20.51$ & $870$   &  $48.01$ \\
123743.08$+$630144.9 & $0.89$ & $10.52$ & $910$   &  $26.82$ \\
130216.13$+$003032.1 & $0.80$ & $3.91$   & $820$   &  $4.47$   \\
130332.42$+$621900.3 & $0.55$ & $4.50$   & $1040$ & $7.18$   \\
133219.66$+$622716.0 & $0.94$ & $7.72$   & $830$   & $12.74$ \\
133422.63$+$475033.6 & $0.93$ & $5.70$   & $950$   & $11.52$ \\
133550.81$+$353315.8 & $0.70$ & $2.95$   & $820$   & $4.03$   \\
135249.82$-$031354.3  & $0.42$ & $9.93$   & $910$   & $13.67$ \\
140300.22$+$432805.4 & $0.85$ & $4.72$   & $990$   & $11.15$ \\
140850.92$+$020522.7 & $1.21$ & $4.48$   & $970$   & $9.36$   \\
144231.72$+$011055.3 & $1.03$ & $2.03$   & $1020$ & $5.21$  \\
153259.96$-$003944.1  & $0.81$ & $4.57$   & $980$   & $6.49$  \\
154734.95$+$444652.5 & $0.67$ & $4.51$   &  $930$   & $4.22$
\enddata
\label{stats}
\tablecomments{The uncertainties on all the best-fit parameters are $\sim15\%$.}
\tablenotetext{a}{Power-law index from least-squares fitting of \hbox{$\lambda f_{\lambda}\propto \lambda^{-\alpha_{\lambda}}$}.}
\tablenotetext{b}{Power-law flux at rest-frame 1~$\mu$m in units of 10$^{-14}$\,erg\,cm$^{-2}$\,s$^{-1}$.}
\tablenotetext{c}{Blackbody temperature.}
\tablenotetext{d}{Blackbody peak flux in units of 10$^{-14}$\,erg\,cm$^{-2}$\,s$^{-1}$.}
\end{deluxetable}

Our results suggest that the two most probable explanations for the weakness of the UV emission lines of WLQs are either
a deficiency of energetic ionizing photons, or a deficiency of line-emitting gas in the BELR.
The first could be a result of a UV-peaked SED (i.e., intrinsic \xray\ weakness) that may be a consequence of an extremely high accretion rate. For example, this may be the case with the high accretion rate quasar PHL~1811 at $z=0.19$ which
is \xray\ weaker by a factor of $\apprge50$ than ordinary quasars with matched optical luminosity (Leighly \et 2007a; Gibson \et 2008).
No WLQ so far observed with sensitive \xray\ imaging is \xray\ weaker than PHL~1811 (although a few WLQs have sensitive \xray\ upper limits and could potentially be weaker; S09; Wu \et 2011).
The peculiar, soft SED of PHL~1811 can naturally explain why the high-ionization emission lines (such as \ion{C}{4}) of the source are unusually weak relative to their low-ionization counterparts
(e.g., \hb; Leighly \et 2007a, 2007b).

However, the extremely high accretion rate scenario for WLQs faces several difficulties.
An extremely high accretion rate would have been expected to result in suppressed IR luminosity with respect to the bolometric luminosity of the accretion disk (e.g., Maiolino \et 2007; Kawakatu \& Ohsuga 2011). We do not have any evidence for significant suppression of IR continuum emission in our WLQs.
In fact, the SEDs of WLQs and ordinary quasars are nearly indistinguishable in the \hbox{$\sim0.1-5~\mu$m} rest-frame band across broad luminosity and redshift ranges for the ordinary quasars; any remaining differences disappear when more comparable luminosity and redshift ranges are considered.
Moreover, the \xray\ properties of at least a subset of our WLQs are similar to the \xray\ properties of ordinary quasars
with comparable luminosities (S09).
It is therefore reasonable to assume that the distributions of the normalized accretion rates (in terms of the Eddington luminosity ratio, \lledd, where $L$ and $L_{\rm Edd}$ represent the bolometric and Eddington luminosities, respectively) of WLQs and ordinary quasars
are similar as well. Recently, Shemmer \et (2010) have directly determined the \lledd\ values for two WLQs from our sample,
SDSS~J1141$+$0219 and SDSS~J1237$+$6301, and found these to be typical of ordinary quasars. In addition, these WLQs
exhibit exceptionally weak \hb\ lines, comparable to the extreme weakness of their UV lines. This result favors the scenario in which the weakness of the WLQ emission lines, across the spectrum, may be attributed to a BELR with abnormal properties, such as a low BELR covering factor, but clearly, no firm conclusions can be drawn based on only two sources.

%%%
%%%
%%%
\begin{deluxetable}{cc}
\tabletypesize{\scriptsize}
\tablecaption{Composite WLQ SED}
\tablewidth{0pt}
\tablehead{ 
\colhead{Central Wavelength } &
\colhead{$\lambda f_{\lambda}$} \\
\colhead{($\mu$m)}	&
\colhead{(Arbitrary Units)}	
}
\startdata
$0.1290^{+0.002}_{-0.002}$  &  1.030  $\pm$  0.087\\
$0.1350^{+0.002}_{-0.002}$  &  1.030  $\pm$  0.085\\
$0.1445^{+0.002}_{-0.002}$  &  1.000  $\pm$ 0.087\\
$0.1695^{+0.002}_{-0.002}$  &  0.947  $\pm$  0.097\\
$0.1830^{+0.002}_{-0.002}$  &  1.030  $\pm$  0.112\\
$0.1965^{+0.002}_{-0.002}$  &  0.967  $\pm$ 0.129\\
$0.2796^{+0.028}_{-0.055}$ & 0.660 $\pm$ 0.052\\
$0.4450^{+0.097}_{-0.083}$ & 0.489 $\pm$ 0.032\\
$0.6400^{+0.069}_{-0.046}$  &  0.319  $\pm$  0.039\\
$0.8200^{+0.078}_{-0.068}$  &  0.250  $\pm$  0.026\\
$0.9930^{+0.055}_{-0.034}$  &  0.203  $\pm$  0.024\\
$1.2300^{+0.152}_{-0.155}$  &  0.199  $\pm$  0.026\\
$1.7100^{+0.414}_{-0.323}$  &  0.244  $\pm$  0.036\\
$4.0600^{+0.235}_{-0.641}$  &  0.409  $\pm$  0.059\\
$5.5200^{+0.878}_{-1.220}$  &  0.335  $\pm$  0.076	
\enddata
\label{mean_sed_table}
\tablecomments{Errors on the central wavelength represent the bin widths; see text for details. Errors on
$\lambda f_{\lambda}$ are the standard deviation of the flux densities in each bin.}
\end{deluxetable}

%%%
%%%
%%%

Another possibility for obtaining a deficiency of high-energy photons, albeit with a typical underlying quasar SED,
is based on a recent investigation of 10 SDSS quasars at
$z\simeq2.2$ selected to have UV-line properties that are similar to those of PHL~1811 (hereafter, `PHL~1811 analogs'; Wu \et 2011).
Wu \et (2011) suggest that the weakness of the UV lines in their PHL~1811 analogs is due to a larger than usual high-ionization ``shielding gas" component that covers most of the BELR, but little more than the BELR in such sources.
In this scenario, such PHL~1811 analogs are a subset of WLQs, viewed through the shielding gas while `normal' WLQs
are viewed at lines-of-sight that do not intersect this component. This may explain the observed \xray\ weakness and apparently
hard \xray\ spectra of the PHL~1811 analogs (Wu \et 2011),
although it is more difficult to explain the very steep (soft) \xray\ spectrum of PHL~1811 itself using this model (Leighly \et 2007a).

Although the evidence from the radio through \xray\ broad-band continua of WLQs as well as the few observations of their rest-frame optical spectra favor the `abnormal BELR' scenario, one cannot draw definitive conclusions about the nature of WLQs without additional
observations. Near-IR spectroscopy of a statistically meaningful sample of WLQs will determine both whether their low-ionization emission lines are as weak as their high-ionization counterparts and whether their accretion rates are exceptionally high. Additional \xray\ spectroscopy of these WLQs can then test the reliability of such \lledd\ determinations if their BELRs are indeed abnormal
(see, e.g., Shemmer \et 2010).
Extended and more detailed coverage of the WLQ SED is also required to allow comparisons between predicted and observed emission-line EWs. For example, the current $\sim0.1-5~\mu$m WLQ SED is only sparsely covered;
considerably denser sampling, in particular in the near-IR band,
would bridge the gap between the available SDSS spectroscopy and \spitzer\ photometry.
Such observations, coupled with photoionization modeling, would improve our understanding of the conditions
necessary for emission-line formation in all AGN.

%%%
%%%
%%%
\begin{deluxetable}{lcc}
\tabletypesize{\scriptsize}
\tablecaption{Richards \et (2006) Quasar Subset}
\tablewidth{0pt}
\tablehead{
\colhead{Quasar} &
\colhead{} &
\colhead{$\log \nu L_{\nu} (1445\,\mbox{\AA})$} \\
\colhead{(SDSS~J)} &
\colhead{$z$} &
\colhead{(erg~s$^{-1}$)} 
}
\startdata
103147.64$+$575858.0 & 2.244 & 46.34\\
103628.15$+$585832.1 & 3.249 & 46.41\\
103750.59$+$590132.1 & 2.126 & 46.20\\
103800.50$+$582343.0 & 3.933 & 46.48\\
103952.54$+$573303.2 & 2.370 & 46.70\\
104018.52$+$572448.1 & 5.215 & 47.30\\
104514.61$+$593707.3 & 4.400 & 46.84\\
104639.42$+$584047.8 & 2.961 & 45.96\\
104809.19$+$570242.0 & 5.132 & 47.64\\
105036.47$+$580424.6 & 2.697 & 46.09\\
105322.98$+$580412.1 & 3.180 & 46.72\\
105524.56$+$580957.3 & 3.064 & 46.24\\
105654.96$+$583712.4 & 2.417 & 46.32\\
105715.49$+$573324.3 & 3.854 & 46.93\\
105902.04$+$580848.7 & 2.139 & 45.85\\
110041.96$+$580001.0 & 3.409 & 45.76\\
161238.27$+$532255.0 & 2.243 & 46.34\\
164022.78$+$411548.1 & 2.879 & 46.33\\
164105.35$+$403651.7 & 2.471 & 45.76\\
164238.08$+$412104.7 & 4.771 & 47.12\\
171652.34$+$590200.1 & 3.123 & 46.54\\
172358.01$+$601140.0 & 2.445 & 46.53
\enddata
\label{rich_subset_props}
\end{deluxetable}

%%%
%%%
%%%
\section{Summary} \label{sec:summary}
%%%
We present new \spitzer\ observations of 14 high-redshift ($z>2.2$) quasars with weak or undetectable emission lines in their rest-frame UV spectra (WLQs) as well as new ground-based, near-IR photometry of two of these sources.
Together with archival \spitzer\ and near-IR photometry as well as rest-frame UV spectroscopy, these observations allowed us to trace the UV-mid-IR SEDs (i.e.,  $\sim0.1-5~\mu$m) of a statistically representative sample of 18 of the \hbox{$\sim80$} known WLQs.
All the SEDs show a clear indication of a typical quasar near-to-mid-IR heated-dust `bump' and they are well represented by a model consisting of a power law plus blackbody; the ranges of best-fit power-law indices ($0.04\leq \alpha_{\lambda}\leq 1.21$) and blackbody temperatures ($820\leq T \leq 1050$) are comparable to the ranges observed in SEDs of ordinary quasars.
We find that all individual SEDs are inconsistent with a relativistically-boosted continuum that might have been expected if WLQs would have been associated with BL Lacertae objects.
By comparing a composite WLQ SED of 17 sources (i.e., excluding SDSS~J1207$+$5953 that has a hint of a \ion{C}{4} BAL trough) with composite SEDs of ordinary quasars with matched luminosities and redshifts, we conclude that WLQs are unbeamed quasars
and that their UV emission lines are intrinsically weak.
The similarity in broad-band continuum properties between WLQs and ordinary quasars, in contrast with the significant differences
between their UV emission-line strengths, 
suggests that WLQs can only be selected efficiently in spectroscopic surveys.
We discuss different scenarios for explaining the weakness of the UV lines in WLQs, mainly extremely high accretion rates or
abnormal broad emission line regions, and describe additional observations required to test both hypotheses.

\acknowledgements

This work is based on observations made with the \spitzer\ Space Telescope, which is operated by the Jet Propulsion Laboratory, California Institute of Technology under a contract with NASA. Support for this work was provided by NASA through an award issued by JPL/Caltech. The work is based in part on observations obtained with the Apache Point Observatory 3.5-meter telescope, which is owned and operated by the Astrophysical Research Consortium. We gratefully acknowledge financial support from the University of North Texas
Research Initiation Grants G34029 and GA9560 (R.~A.~L, O.~S), from the Southern California Center for Galaxy
Evolution, a multi-campus research program funded by the University of California Office of Research (A.~M.~D),
from a NASA ADP grant NNX10AC99G (W.~N.~B), and from an NSF grant AST-0707266 (M.~A.~S).
We thank an anonymous referee for a helpful report, and the members of the \spitzer\ helpdesk team for assistance with the data reduction.
This research has made use of the NASA/IPAC Extragalactic Database (NED) which is operated by the Jet propulsion Laboratory, California Institute of Technology, under contract with the National Aeronautics and Space Administration.
Funding for the SDSS and SDSS-II has been provided by the Alfred P. Sloan Foundation, the Participating Institutions, the National Science Foundation, the U.S. Department of Energy, the National Aeronautics and Space Administration, the Japanese Monbukagakusho, the Max Planck Society, and the Higher Education Funding Council for England. The SDSS Web Site is http://www.sdss.org/.

{{\it Facilities:} \facility{\sl Spitzer ()}, \facility{Sloan()}}, \facility{ARC ()}
%%%
%%%
%%%
%%%

%%%%
%%%%
%%%%
%%%%

\begin{thebibliography}{}
%
\bibitem[Anderson et al.(2001)]{2001AJ....122..503A} Anderson, S.~F., et 
al.\ 2001, \aj, 122, 503 
%
\bibitem[Becker et al.(1995)]{1995ApJ...450..559B} Becker, R.~H., White, 
R.~L., \& Helfand, D.~J.\ 1995, \apj, 450, 559
%
\bibitem[Cardelli et al.(1989)]{1989ApJ...345..245C} Cardelli, J.~A., 
Clayton, G.~C., \& Mathis, J.~S.\ 1989, \apj, 345, 245
%
\bibitem[Collinge et al.(2005)]{2005AJ....129.2542C} Collinge, M.~J., et 
al.\ 2005, \aj, 129, 2542 
%
\bibitem[Condon et al.(1998)]{1998AJ....115.1693C} Condon, J.~J., Cotton, 
W.~D., Greisen, E.~W., Yin, Q.~F., Perley, R.~A., Taylor, G.~B., 
\& Broderick, J.~J.\ 1998, \aj, 115, 1693
%
\bibitem[Diamond-Stanic et al.(2009)]{2009ApJ...699..782D} Diamond-Stanic, 
A.~M., et al.\ 2009, \apj, 699, 782 (DS09)
%
\bibitem[Elvis et al.(1994)]{1994ApJS...95....1E} Elvis, M., et al.\ 1994, 
\apjs, 95, 1
%
\bibitem[Fan et al.(1999)]{1999ApJ...526L..57F} Fan, X., et al.\ 1999, 
\apjl, 526, L57 
%
\bibitem[Fan et al.(2001)]{2001AJ....121...31F} Fan, X., et al.\ 2001, \aj, 
121, 31
%
\bibitem[Fan et al.(2006)]{2006AJ....131.1203F} Fan, X., et al.\ 2006, \aj, 
131, 1203
%
\bibitem[Fazio et al.(2004)]{2004ApJS..154...10F} Fazio, G.~G., et al.\ 
2004, \apjs, 154, 10
%
\bibitem[Fossati et al.(1998)]{1998MNRAS.299..433F} Fossati, G., Maraschi, 
L., Celotti, A., Comastri, A., \& Ghisellini, G.\ 1998, \mnras, 299, 433 
%
\bibitem[Gibson et al.(2008)]{2008ApJ...685..773G} Gibson, R.~R., Brandt, 
W.~N., \& Schneider, D.~P.\ 2008, \apj, 685, 773
%
\bibitem[Gibson et al.(2009)]{2009ApJ...692..758G} Gibson, R.~R., et al.\ 
2009, \apj, 692, 758
%
\bibitem[Hryniewicz et al.(2010)]{2010MNRAS.404.2028H} Hryniewicz, K., 
Czerny, B., Niko{\l}ajuk, M., \& Kuraszkiewicz, J.\ 2010, \mnras, 404, 2028 
%
\bibitem[Jannuzi et al.(1993)]{1993ApJ...404..100J} Jannuzi, B.~T., Green, 
R.~F., \& French, H.\ 1993, \apj, 404, 100 
%
\bibitem[Jiang et al.(2006)]{2006AJ....132.2127J} Jiang, L., et al.\ 2006, 
\aj, 132, 2127
%
\bibitem[Jiang et al.(2010)]{2010Natur.464..380J} Jiang, L., et al.\ 2010, 
\nat, 464, 380
%
\bibitem[Kaspi et al.(2007)]{2007ApJ...659..997K} Kaspi, S., Brandt, W.~N., 
Maoz, D., Netzer, H., Schneider, D.~P., 
\& Shemmer, O.\ 2007, \apj, 659, 997
%
\bibitem[Kawakatu \& Ohsuga(2011)]{2011arXiv1107.2185K} Kawakatu, N., \& Ohsuga, K.\ 2011, arXiv:1107.2185 
%
\bibitem[Kellermann et al.(1989)]{1989AJ.....98.1195K} Kellermann, K.~I., 
Sramek, R., Schmidt, M., Shaffer, D.~B., \& Green, R.\ 1989, \aj, 98, 1195 
%
\bibitem[Leighly et al.(2007)]{2007ApJ...663..103L} Leighly, K.~M., 
Halpern, J.~P., Jenkins, E.~B., Grupe, D., Choi, J., 
\& Prescott, K.~B.\ 2007a, \apj, 663, 103 
%
\bibitem[Leighly et al.(2007)]{2007ApJS..173....1L} Leighly, K.~M., 
Halpern, J.~P., Jenkins, E.~B., \& Casebeer, D.\ 2007b, \apjs, 173, 1 
%
\bibitem[Liu 
\& Zhang(2011)]{2011ApJ...728L..44L} Liu, Y., \& Zhang, S.~N.\ 2011, \apjl, 728, L44 
%
\bibitem[Londish et al.(2004)]{2004MNRAS.352..903L} Londish, D., Heidt, J., 
Boyle, B.~J., Croom, S.~M., 
\& Kedziora-Chudczer, L.\ 2004, \mnras, 352, 903 
%
\bibitem[Makovoz 
\& Marleau(2005)]{2005PASP..117.1113M} Makovoz, D., \& Marleau, F.~R.\ 2005, \pasp, 117, 1113 
%
\bibitem[Maiolino et al.(2007)]{2007A&A...468..979M} Maiolino, R., Shemmer, O., Imanishi, M.,
Netzer, H., Oliva, E., Lutz, D., \& Sturm, E.\ 2007, \aap, 468, 979 
%
\bibitem[McDowell et al.(1995)]{1995ApJ...450..585M} McDowell, J.~C., 
Canizares, C., Elvis, M., Lawrence, A., Markoff, S., Mathur, S., 
\& Wilkes, B.~J.\ 1995, \apj, 450, 585
%
\bibitem[Nieppola et 
al.(2006)]{2006A&A...445..441N} Nieppola, E., Tornikoski, M., \& Valtaoja, E.\ 2006, \aap, 445, 441 
%
\bibitem[Oke et al.(1995)]{1995PASP..107..375O} Oke, J.~B., et al.\ 1995, 
\pasp, 107, 375
%
\bibitem[Padovani 
\& Giommi(1995)]{1995ApJ...444..567P} Padovani, P., \& Giommi, P.\ 1995, \apj, 444, 567 
%
\bibitem[Plotkin et al.(2010)]{2010AJ....139..390P} Plotkin, R.~M., et al.\ 
2010a, \aj, 139, 390 
%
\bibitem[Plotkin et al.(2010)]{2010ApJ...721..562P} Plotkin, R.~M., 
Anderson, S.~F., Brandt, W.~N., Diamond-Stanic, A.~M., Fan, X., MacLeod, 
C.~L., Schneider, D.~P., \& Shemmer, O.\ 2010b, \apj, 721, 562 
%
\bibitem[Richards et al.(2006)]{2006ApJS..166..470R} Richards, G.~T., et 
al.\ 2006, \apjs, 166, 470 (R06)
%
\bibitem[Rieke et al.(2004)]{2004ApJS..154...25R} Rieke, G.~H., et al.\ 
2004, \apjs, 154, 25
%
\bibitem[Schneider et al.(2005)]{2005AJ....130..367S} Schneider, D.~P., et 
al.\ 2005, \aj, 130, 367 
%
\bibitem[Schlegel et al.(1998)]{1998ApJ...500..525S} Schlegel, D.~J., 
Finkbeiner, D.~P., \& Davis, M.\ 1998, \apj, 500, 525 
%
\bibitem[Shemmer et al.(2006)]{2006ApJ...644...86S} Shemmer, O., et al.\ 
2006, \apj, 644, 86 (S06)
%
\bibitem[Shemmer et al.(2009)]{2009ApJ...696..580S} Shemmer, O., Brandt, 
W.~N., Anderson, S.~F., Diamond-Stanic, A.~M., Fan, X., Richards, G.~T., 
Schneider, D.~P., \& Strauss, M.~A.\ 2009, \apj, 696, 580 (S09)
%
\bibitem[Shemmer et al.(2010)]{2010ApJ...722L.152S} Shemmer, O., et al.\ 
2010, \apjl, 722, L152
%
\bibitem[Skrutskie et al.(2006)]{2006AJ....131.1163S} Skrutskie, M.~F., et 
al.\ 2006, \aj, 131, 1163
%
\bibitem[Stocke 
\& Perrenod(1981)]{1981ApJ...245..375S} Stocke, J.~T., \& Perrenod, S.~C.\ 1981, \apj, 245, 375 
%
\bibitem[Stocke et al.(1990)]{1990ApJ...348..141S} Stocke, J.~T., Morris, 
S.~L., Gioia, I., Maccacaro, T., Schild, R.~E., 
\& Wolter, A.\ 1990, \apj, 348, 141
%
\bibitem[Trump et al.(2006)]{2006ApJS..165....1T} Trump, J.~R., et al.\ 
2006, \apjs, 165, 1
%
\bibitem[Urry 
\& Padovani(1995)]{1995PASP..107..803U} Urry, C.~M., \& Padovani, P.\ 1995, \pasp, 107, 803 
%
\bibitem[Vanden Berk et al.(2001)]{2001AJ....122..549V} Vanden Berk, D.~E., 
et al.\ 2001, \aj, 122, 549 
%
\bibitem[Werner et al.(2004)]{2004ApJS..154....1W} Werner, M.~W., et al.\ 
2004, \apjs, 154, 1
%
\bibitem[Wu et al.(2011)]{2011ApJ...736...28W} Wu, J., et al.\ 2011, \apj, 
736, 28
%
 \bibitem[York et al.(2000)]{2000AJ....120.1579Y} York, D.~G., et al.\ 2000, 
\aj, 120, 1579 

\end{thebibliography}
\end{document}